# An Android App for Digital Transport Cards of Smart City: Request for Suggestions

Qiaoli Wang     414217795zoe@gmail.com

*Abstract: This paper provides a roadmap for designing the Global Digital Transport Cards Application (GDTCA). GDTCA is an Android Application for people to get a digital transport card online for any city and use it locally. People can top up or withdraw the transport card by using GDT Cards App. It is worth noting that such an application will have different local learning experiences based on the sparsity of the environments, which can be aggregated using federated learning and transport protocols.*

Introduction

Market Research and Motivation

A news was published on 9news that when the Myki card is lost or expired, the current system does not allow the card's remaining funds to be transferred to the new card. Due to the lost or expired Myki pass, nearly 80 million US dollars of discarded funds have been lost [1]. So I aim to solve the problem about withdraw funds from transportation card and get transportation cards more conveniently.

The main problem with getting a transportation card more conveniently is that there is no online way to register a digital transportation card.  So, I will develop an application for apply for digital transportation cards online. Which is Global Digital Transportation Cards, I will be called it GDT cards app.

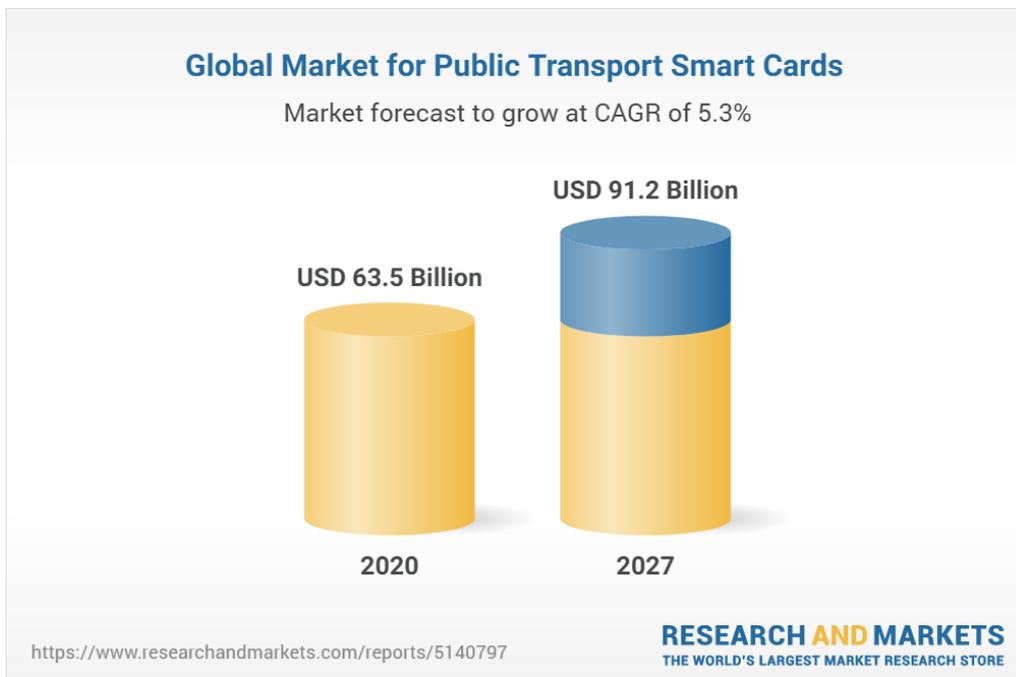

*Figure 1 Global market for public Transport Smart Cards*

## Background and Summary of the intended product.

People use public transport every day. Over 12 million Australians aged 14+ (58%) use public transport in an average of three months [2]. how people can buy and manage transportation cards more conveniently to prevent the cards from being lost or stolen.

The GDT cards app allows users to register digital transportation cards online as well as manage and recharge them online for any city and use it locally. Now there is close to $80 million in discarded funds on lost or expired Myki passes. Assume we can manage transportation card online, recharge, withdraw, remove the card. Service will faster and more reliable, which has led to an increase in commuters. both greenhouse gas and carbon emissions can be reduced.

## Competitor Analysis

The current competitors for digital transportation cards include Google Pay, Apple Pay, and Samsung Pay. These are mobile phone wallets. These competitors only can add Physical card according to the different transportation system, GDT Cards app is meant to apply for a new digital card online and people can use it locally and manage it online.

## Features

### Asset list

| Asset Name | Statistical Details | Main Components |
|---|---|---|
| Get start page | 1.Image to display the logo of the GDT card app.<br><br>2. TextView for display information about introducing the app.<br><br>3.Button for the user to click to navigate to Register or login page. | Buttons, Image, TextView |
| Register page | 1.Text fields let users enter and edit their information (eg. email, password etc).<br><br>2.Buttons let user can click to register. | Text fields, Buttons |
| Login page | 1.Text fields let users enter and edit text.<br><br>2.Checkboxes can turn to remember password on or off. | Text fields, Buttons, Checkboxes |

|  | 3. Button for the user to click to login |  |
|---|---|---|
| Home page(cards list page) | 1.Cards contain content and actions about a single subject. User can click card item to navigate to the card details page.<br><br>2.A floating action button let the user click to Apply for a new card. | Cards, A floating action button |
| Apply new card page | 1.Text fields let users enter and edit text.<br><br>2.Radio buttons allow users to select one option from a set.<br><br>3. button allows the user to click to submit this application. | Text fields, Buttons, Radio buttons |
| Card details page | 1.Bar code API can generate a number to a bar code<br><br>2.Buttons on the card details page allow the user to recharge the card or withdraw funds from the card. | Bar code API, Buttons, |
| Wallet (Transactions history page) | 1.Bar code API for the cards display with a bar code<br><br>2.Line graph for the data visualization for transaction history<br><br>3. Top app bar for navigation back<br><br>4. use ViewPager to create a Screen slides are transitions between one card to another | Bar code API, Line graph, Top app bar for navigation back, ViewPager |
| Transaction details page | 1.Button for click to back to home page<br><br>2. TextView for showing detailed information. | TextView, Button |
| User profile page | 1.when the user click "forget password" it will display a Dialog for change the password<br><br>2.Text fields let users enter and edit text.<br><br>3.Button for submit to update information | Text fields, Buttons, Dialog |
| Payment page | 1.Text fields let users enter and edit text.<br><br>2.Radio buttons allow users to select one amount from a set.<br><br>3.Button for click to next step | Text fields, Buttons, Radio buttons |
| Confirm payment page | 1.Text fields allow the user to enter the note<br><br>2.Button for the user to click to next step | Text fields, Buttons |
| Payment receipt | 1. TextView for showing the payment information.<br><br>2.Button for the user to click back to the home page. | TextView, Buttons |
| Withdraw page | 1.Dialog for when user click "withdraw all"<br><br>2.Text fields here is to display all funds of the card.<br><br>3.Checkboxes for when people select to withdraw all and want to remove the card | Text fields, Buttons, Dialog, Checkboxes |
| Confirm withdraw page | 1.Text fields allow the user to enter the note.<br><br>2.Button for the user to click to next step | Text fields, Buttons |
| Withdraw receipt page | 1. Button for the user to click to back home page<br><br>2. TextView for showing Withdraw information. | Buttons, TextView |

Product purpose

**Target audience**

The target audience of the GDT Cards APP is the groups who use public transportation for daily travel, as well as the users who like to use public transportation for the first time to a new city. the GDT Cards APP can provide passengers who use public transportation daily to obtain transportation cards more conveniently, recharge transportation cards, and withdraw the balance in the card when they stop using a certain card. and when you travel to a new city, you can register a digital transportation card online, which can be used locally.

**The opportunities and benefits of my project.**

- o  There are significant cost savings associated with going digital.
- o  People can apply for online transportation cards for any city in the world.
- o  Fast and easy payment for passengers eliminates the need to queue for a paper ticket or to 'top-up' smart tickets.
- o  People travel by public transportation will become smarter and easier by using digital transportation card, and the transportation card will not be lost.

## 3 Complex components

For our project, we will use Kotlin Android to develop our APP.

**Component 1: Card details page**

On the card details page, we should display a barcode to allow the user to scan when using public transport (we assume public transport have a barcode reader).

We need to generate a barcode in Kotlin Android.

1. Add ZXing library
   - add a dependency on the ZXing core library

2. Set up a dimensions file
   - it's useful to create a dimensions file to hold the size of your barcode image which can be accessed from the layout and the code.

3. Create the layout
   - Replace the default TextView with an ImageView to hold the barcode and a TextView below it to hold the barcode value.

4. Edit MainActivity.kt
   - Add a function to generate the barcode image
   - Add a helper method that sets the size of the barcode image and uses the previous function to set the actual barcode image, it also sets the text value
   - Finally, call the helper method from OnCreate.

**Component 2: Apply a new transportation card page**

On this page, we should allow the user to pay for her/his transportation card. So we should allow the user to add a payment card. The complex of this is that how we can verify the payment card and complete payment.

I will create a credit card form first. if verity and payment with add cards are hard, I will Integrate Google Pay into the application.

1. Add Google Play services to my build. Gradle
2. Enable the API in my manifest file
3. Plan my user interface and where to show the Google Pay button
4. Initialize and configure the Google Pay API
5. Determine readiness to pay with Google Pay

**Component 3: Wallet page**

On the wallet page, we should display a line chart for the user to look at the transaction history for each month. User can click the line point to check detail.

So we need to add an android library about line chart.

I didn't decide to use which library for my line chart. but I will introduce the step to use a line chart library with android.

1. Add the repository to my build file
2. Add the dependency
3. we'll need a reference to a view first, either through XML or programmatically:
4. Change the style
5. Define a dataset for the line chart to display the data.

## Summary of Project plan

The development of the system will divide into 3 sprints. The following is about a task for each sprint and the due day of each task. The following picture shows that we intend to complete the development of this APP by end of May.

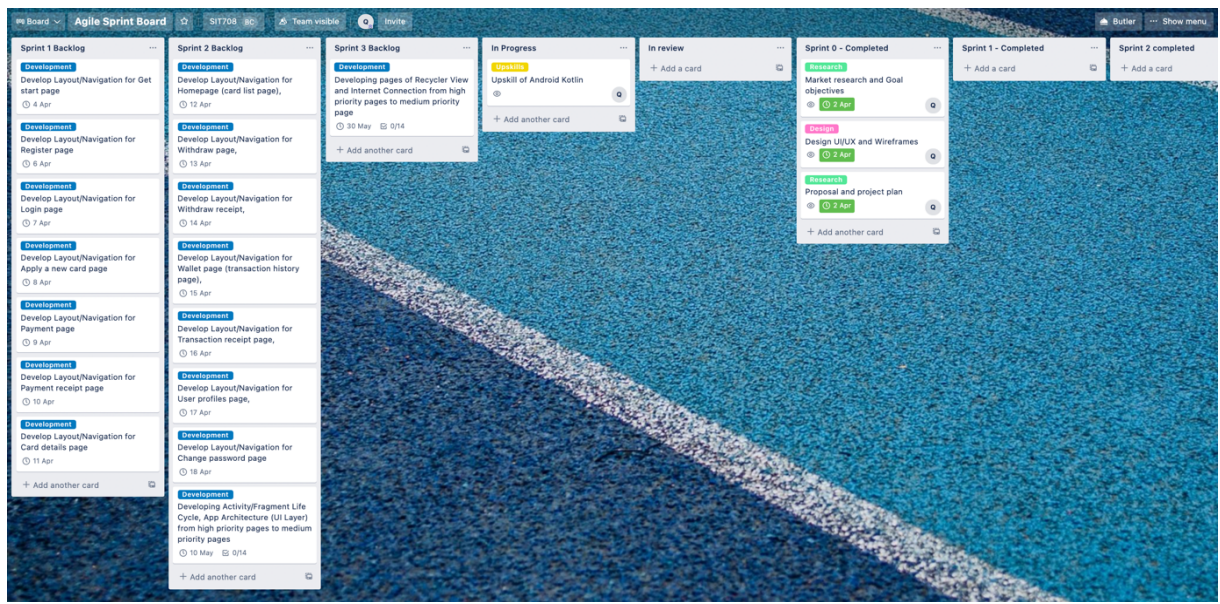

*Figure 2 screenshot of Trello board*

## Milestones of the project plan

**Sprint 0 (week 1 – week 4)**
- Market research and Goal objectives
- Design UI/UX and Wireframes
- Proposal and project plan
- Upskill of Android Kotlin

**Sprint 1(Week5 – week6)**
- Developing Android Apps, Layout/Navigation from height priority pages to medium priority pages

1. Get start page
2. Register page
3. Login page
4. Apply a new card page
5. Payment page
6. Payment receipt page
7. Card details page

**Sprint 2 (Week 6 – Week 7)**
- Develop Layout/Navigation for remaining medium priority pages

1. Homepage (card list page),
2. Withdraw page,
3. Withdraw receipt,
4. Wallet page (transaction history page),
5. Transaction receipt page,
6. User profiles page,
7. Change password page

- Developing Activity/Fragment Life Cycle, App Architecture (UI Layer) from high priority pages to medium priority pages

**Sprint 3 (week 8 – week 9)**
- Developing Activity/Fragment Life Cycle, App Architecture (UI Layer) from high priority pages to medium priority pages

- Developing pages of Recycler View and Internet Connection from high priority pages to medium priority page

## Detailed UX/UI Design

### User Stories
**User Story: 1**

| Statement | Acceptance Criteria | Estimation | Priority |
|---|---|---|---|
| As a user, I want to apply for a transportation card directly on my mobile phone, so that I can use it when I use public transport. | 1.User should able to create an account for this app.<br><br>2.User should able to login into the app.<br><br>3. User should able to apply for a transportation card on the app<br><br>4. It should have a card details page showing a transportation card with the card number and bar code. | According to the planning poker estimate approach, this can be considered as<br><br>**Story Points: 12** | **Priority: 1**<br><br>High Priority<br><br>**Prerequisite of User Story 2,4 and 5** |

**User Story: 2**

| Statement | Acceptance Criteria | Estimation | Priority |
|---|---|---|---|
| As a traveller, I want to register a transportation card directly on my phone for my next destination, so that I can use it when I arrive destination. | 1.User should able to enter or select a city that he/she wants to apply for the card. | According to the planning poker estimate approach, this can be considered as<br><br>**Story Points: 1** | **Priority: 1**<br><br>High Priority |

**User Story: 3**

| Statement | Acceptance Criteria | Estimation | Priority |
|---|---|---|---|
| As a user, I want to check my transportation card on my phone, so that I can track my transaction and purchase. | 1.User should able to view all transportation cards.<br><br>2. User should able to check the transaction and purchase history. | According to the planning poker estimate approach, this can be considered as<br>**Story Points: 6** | **Priority: 1**<br><br>High Priority |

**User Story: 4**

| Statement | Acceptance Criteria | Estimation | Priority |
|---|---|---|---|
| As a user, I want to be able to directly recharge or pay for my transportation on my phone, so that I don't need to recharge it at a machine. | 1. User should able to view the card details<br><br>2. should have a recharge button on the card details page.<br><br>3.when click the "recharge" button, the user should able to select or enter an amount that the user wants to recharge.<br><br>3. User should able to add a payment card to pay for it.<br><br>4. user should able to check if the payment is successful or not. | According to the planning poker estimate approach, this can be considered as<br><br>**Story Points: 10** | **Priority: 1**<br><br>High Priority |

**User Story: 5**

| Statement | Acceptance Criteria | Estimation | Priority |
|---|---|---|---|
|  |  |  |  |

| Statement | Acceptance Criteria | Estimation | Priority |
|---|---|---|---|
| As a user, I want to withdraw the card amount, so that I won't waste the balance on the card when I don't use it. | 1. It should have a "withdraw" button on the card details page.<br><br>2.User should able to enter an amount that he/she wants to withdraw.<br><br>3. User should able to see the "withdraw receipt page" to check if withdraw successfully or not. | According to the planning poker estimate approach, this can be considered as<br><br>**Story Points: 6** | **Priority: 1**<br><br>High Priority<br><br>**Prerequisite of User Story 6** |

**User Story: 6**

| Statement | Acceptance Criteria | Estimation | Priority |
|---|---|---|---|
| As a user, I want to remove a transport card when I intend not to use it so that this card can also be released for other people to use. | 1.it should have a withdraw all button on the withdraw page.<br><br>2.when the user clicks "withdraw all", it should have a checkbox for the user to choose "remove card" or not. | According to the planning poker estimate approach, this can be considered as<br><br>**Story Points: 4** | **Priority: 2**<br><br>High Priority |

**User Story: 7**

| Statement | Acceptance Criteria | Estimation | Priority |
|---|---|---|---|
| As a user, I want to see the summary of my transaction for each transportation card, so that I can check the money I spent on public transport for each month. | 1.it should have a line graph on the wallet page.<br><br>2. User should able to click each line to check the amount for each month. | According to the planning poker estimate approach, this can be considered as<br><br>**Story Points: 6** | **Priority: 3**<br><br>Medium Priority |

**User Story: 8**

| Statement | Acceptance Criteria | Estimation | Priority |
|---|---|---|---|
| As a user, I want to update my information and change my password, so that I can keep my information up-to-date. | 1.it should have a user profile page that allows the user to update information.<br><br>2. User should able to click the "change password" button to change the password. | According to the planning poker estimate approach, this can be considered as<br><br>**Story Points: 4** | **Priority: 3**<br><br>Medium Priority |

## User Cases

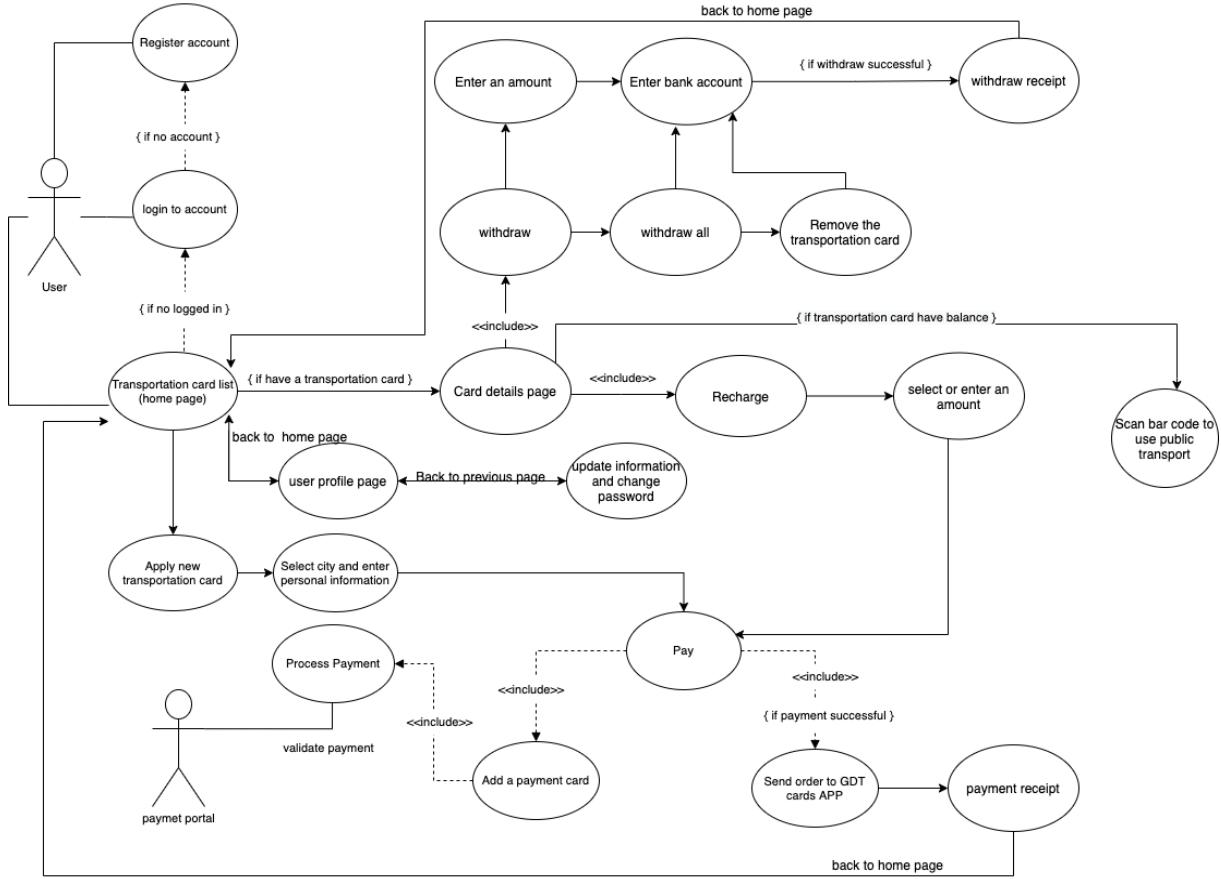

*Figure 2 User case*

URL to UX/UI Design

https://www.figma.com/proto/o5d7wP4snrK2vGG1mXm5Nh/GDT-Cards---Register-transport-card-online?node-id=212%3A1&scaling=scale-down&page-id=0%3A1

## High-Level Wireframes

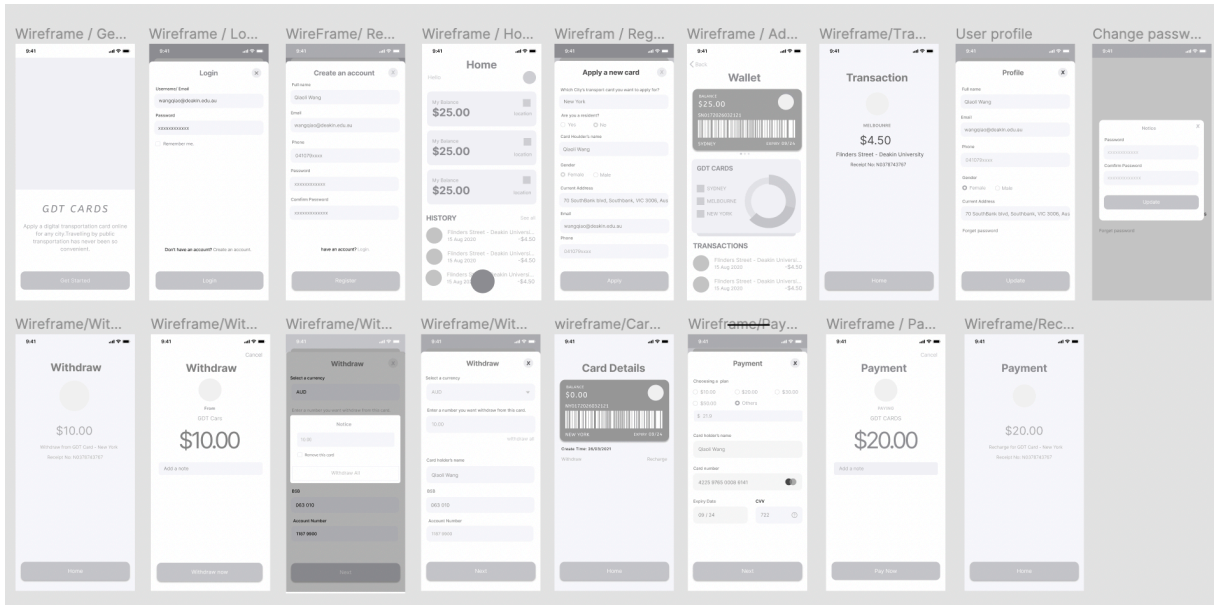

*Figure 3 wireframes*

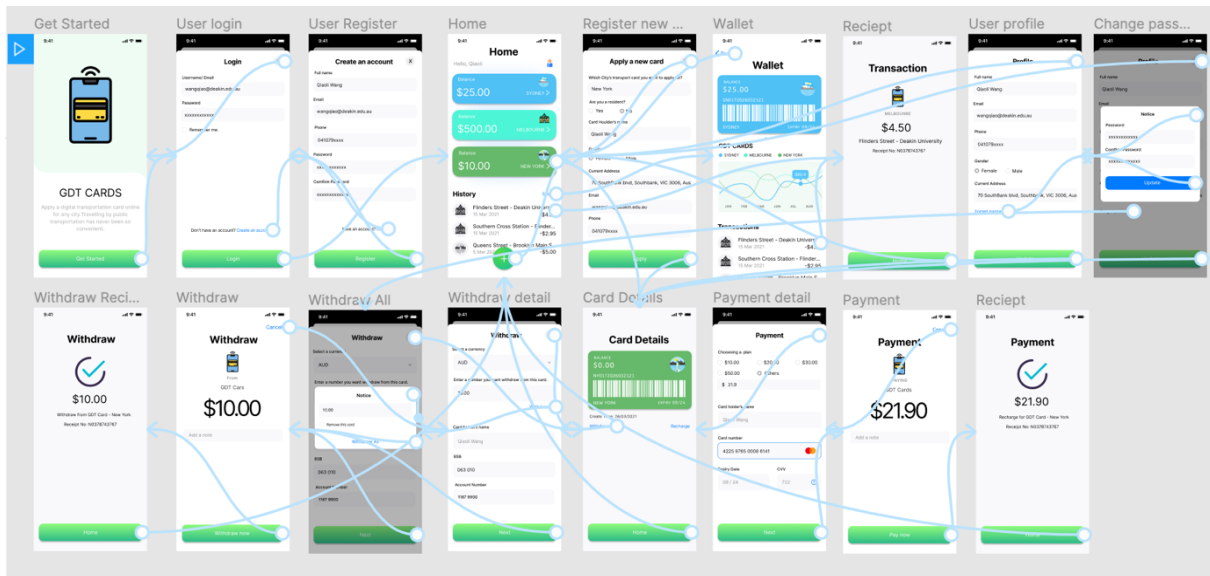

*Figure 4 prototype of UI/UX design*

## Resources required

URL of Wireframe:
https://www.figma.com/file/o5d7wP4snrK2vGG1mXm5Nh/GDT-Cards---Register-transport-card-online?node-id=274%3A55

Prototype of UI/UX Design:
https://www.figma.com/file/o5d7wP4snrK2vGG1mXm5Nh/GDT-Cards---Register-transport-card-online?node-id=0%3A1

Link of Trello Board:
https://trello.com/invite/b/Ct9inlFD/de0f2d457b5610d95f2936397c0fb963/agile-sprint-board

GitHub Link: https://github.com/Qiaoliwang2020/GDTCards-Android-App

## Performance Evaluation and Future Directions

In this work, we will perform rigorous testing of the developed application over transport layer protocols including multipath TCP [3], and TCP [4] over the widely used last-miles 802.11 and 4G networks [5]. For future advances in the apps, we seek ways to adapt to intelligent public transportation systems in distributed computing systems of autonomous vehicles [6].